\documentclass{article}

\usepackage{graphics}
\usepackage[english]{babel}   
\usepackage{graphicx}
\usepackage{amsmath}
\usepackage{amssymb}
\usepackage{multicol}
\usepackage{multirow}
\usepackage{tabularx}
\usepackage{xcolor}
\newcolumntype{C}[1]{>{\centering\let\newline\\\arraybackslash\hspace{0pt}}m{#1}}
\def\PP{\mathbb{P}}

\usepackage[colorlinks=true,linkcolor=blue]{hyperref}
\usepackage[normalem]{ulem}

\begin{document}

\title{Probabilistic risk bounds for the characterization of radiological contamination}
\author{G\'eraud Blatman$^\star$, Thibault Delage$^\dag$, Bertrand Iooss$^\dag$  and Nadia P\'erot$^\ddag$\\
$^\star$ EDF Lab Les Renardi\`eres, 77818 Moret-sur-Loing, France\\
$^\dag$ EDF Lab Chatou, 78401 Chatou, France\\
$^\ddag$ CEA Nuclear Energy Division, 13108 Saint-Paul-l\` es-Durance, France
}
% \thanks is optional - remove next line if not needed
%\thanks{\emph{Present address:} Insert the address here if needed}%
                     % Do not remove
%
%\offprints{}          % Insert a name or remove this line
%
%
% The correct dates will be entered by Springer
%
\maketitle

\abstract{
The radiological characterization of contaminated elements (walls, grounds, objects) from nuclear facilities often suffers from too few measurements. 
In order to determine risk prediction bounds on the level of contamination, 
some classic statistical methods may therefore be unsuitable, as they rely upon 
strong assumptions (e.g., that the underlying distribution is Gaussian) which 
cannot be verified. Considering that a set of measurements or their average 
value come from a Gaussian distribution can sometimes lead to erroneous 
conclusions, possibly not sufficiently conservative. 
This paper presents several alternative statistical approaches which are 
based on much weaker hypotheses than the Gaussian one, which result from 
general probabilistic inequalities and order-statistic based formulas. 
Given a data sample, these inequalities make it possible to derive prediction intervals for a random variable which can be directly interpreted as probabilistic risk bounds. 
For the sake of validation, they are first applied to simulated data  generated from several known theoretical distributions. 
Then, the proposed methods are applied to two data sets obtained from real radiological contamination measurements. 
%

%\PACS{
%      {PACS-key}{discribing text of that key}   \and
%      {PACS-key}{discribing text of that key}
%     } % end of PACS codes
} %end of abstract

%\keywords{Confidence interval, Nuclear waste, Prediction, Probabilistic inequality, Radiological contamination, Sampling, Wilks Formula}
%
%\tableofcontents
%

\section{Introduction}

In nuclear engineering, as for most industrial domains, one often faces a difficult decision-making processes, especially when safety issues are involved. In order to consider in a rigorous and consistent way all environment uncertainties in a decision process, a probabilistic framework offers invaluable help. For example, typical 
non-exhaustive sampling of a process or object induces uncertainty that needs to be understood in order to control its effects. 

In particular, the estimation of risk prediction bounds is an important element of a comprehensive probabilistic risk assessment of radioactive elements (e.g., walls, grounds, objects) derived from the nuclear industry.
The radiological characterization of contaminated elements in a nuclear facility may be difficult because of practical and/or strong operation constraints, often limiting the number of possible measurements. 
Nevertheless, the estimation of  radioactivity levels is essential to assess the risk of exposure of  nuclear dismantling operators, as well as the risk of environment contamination~\cite{cet14}.

Performing a realistic and reasonable risk estimate is essential, not only from an economic perspective, but also for public acceptance. First, this is important information for decision makers in order to be able to implement suitable measures that are financially acceptable. Nevertheless, overestimating the risk often turns out to be counterproductive by inducing much uncertainty and a loss of confidence in public authorities from the population. We cite for example the management of the 2009 flu epidemic (H1N1), where many countries' authorities had a disproportionate reaction with regard to real risk, due to the World Health Organization's (WHO) poorly managed forecasts~\cite{WHO2009}. This  led to public loss of confidence in the authorities, in vaccines, and in pharmaceutical companies, as well as an enormous waste of public resources.

%A statistical problem strongly arises when the quantification of a radionuclide inventory can be only made using a small number of measurements (of the order of $10$).
Drawing up a radiological inventory based on a small number of measurements 
(e.g., in the order of 10) is a particularly difficult statistical problem.  
The shortage of data can lead either to a coarse over-estimation, which has large impact on economic cost, or to a coarse under-estimation, which has an unacceptable impact in terms of public health and environment protection.
In the past, several attempts have been made to deal with such problems. For instance,
\cite{perioo08} focused on the problem of defining a sampling strategy and assessing the representativeness of the small samples at hand.
In the context of irradiated graphite waste, \cite{ponpet13} developed a method to assess the radionuclide inventory as precisely as possible with a $2.5$\% risk of under-assessment.
In a recent work, \cite{zafmag16} described several sampling methods to estimate the concentration of radionuclides in radioactive waste, by using correlations between different radionuclide's activities.
When the  contamination characterized exhibits a certain spatial continuity and when the spatial localization of  measurements can be chosen, geostatistical tools can be used, as shown in \cite{jeades08,deschi11,becrom13}.

In this work, we focus on the difficult task of radiological characterization based on a small number of data which are assumed  statistically independent (and non-spatially localized). 
This task belongs to a quite general class of problems: the statistical analysis of small data samples (see for example \cite{bradav07,puppup10,schneu09}).
In this case,  classical statistical tools turn out to be unsuitable. 
For example, assuming that a set of measurements or their average value arise from a Gaussian distribution can lead to erroneous and sometimes non-conservative conclusions. 
Indeed, if the estimation of the mean value is of interest, Gaussian distribution-based bounds may only be used in the asymptotic limit of a very large sample, and the convergence to this asymptotic regime may be very slow in the presence of a noticeably skewed actual data-generating distribution.
Even if some solutions exist to correct this large-size sample requirement, the Gaussian distribution hypothesis may still be invalid or impossible to justify by rigorous statistical tests \cite{dagste86}. 

Alternative statistical tools, called concentration inequalities (but also denoted universal inequalities or robust inequalities), are applicable without knowing the probability distribution of the  variable being studied. 
In general, from a data sample, statistics-based intervals  allow the derivation of \cite{hahmee91}:
\begin{enumerate}
\item	Confidence intervals for the estimation of the mean (or other distribution parameters) of a random variable. For example, we can determine the size of the set of measurements  to make in order to reach a given precision for calculating the average value of various contamination measures.
This process allows us to optimize the sampling strategy and offers invaluable economic gains.
\item	Prediction intervals for a random variable. For example, we can compute the probability that the value of a point contamination is larger than a given critical value. In practice, regulatory threshold values are set for different waste categories.
Determination of the probability that the contaminant's value is smaller than a given threshold can be used to predefine the volumes of waste by category.
\item Tolerance intervals, which extend prediction intervals to take into account  uncertainty in the parameters of a variable's distribution.
A tolerance interval gives the statistical interval within which, with some confidence level, a specified proportion of a sampled population falls. 
\end{enumerate}

In this paper, we focus on the second and third intervals mentioned above (note that confidence intervals are also addressed in \cite{blaioo12}).
%, using the term ``prediction intervals'' in the following for the sake of simplicity.
Easy to state and easy to use, the Bienaym\'e-Chebychev inequality \cite{nel95} is the most famous probabilistic inequality.
Unfortunately, it comes at the expense of extremely loose bounds, which make it unsuitable in  practical situations, so it is not  used often.
From these considerations, \cite{woo88} has proposed to use the more efficient (but little-known) Guttman inequality \cite{gut48a}.
Even if it requires no additional assumptions, it has however the drawback of requiring an estimate of the kurtosis (i.e., the fourth-order statistical moment) of the  variable being studied.
In the small dataset context  (around ten points), a precise estimation of the kurtosis might seem to be an unrealistic goal.

In another context, that of the quality control domain, \cite{puk94} has developed narrower bounds than the Bienaym\'e-Chebychev ones, showing at the same time how the three-sigma rule can be justified (based on a unimodality assumption, proven for example in \cite{vyspet80}).
Starting with this statistical literature as a base,  along with older results about unimodal and convex distributions (see \cite{dhajoa88} for a more recent example), several useful inequalities have been developed in \cite{blaioo12}.
While they also focused on the range of validity  of each inequality, their results were preliminary; the present paper extends that work to estimate risk prediction bounds robustly, with several applications to real radiological characterization problems. 
Furthermore, we make a connection between this risk bound estimation problem and the problem of computing conservative estimates of a quantile, classically addressed in nuclear thermal-hydraulic safety using the so-called Wilks formula \cite{nutwal04}.
Comparisons are then performed between the various approaches.

The following section provides all of the probabilistic inequalities that we can use to attempt to solve our problems.
For validation purposes, all of these are applied in Section 3 to simulated data samples generated from several known theoretical distributions. 
More specifically, the accuracy of the resulting prediction and tolerance intervals are compared to those obtained from standard methods such as the Gaussian approximation. 
Section 4 shows how the probabilistic inequalities can be used in practice, and more precisely, to analyze radiological contamination measures.
A conclusion follows to summarize the results of this work.

%%%%%%%%%%%%%%%%%%%%%%%%%%%%%%%%%%%%%%%%%%%%%%%%%%%%%%%%%%%%%%%%%%%%%%%%
\section{Probabilistic inequalities for prediction and tolerance intervals}

%%%%%%
%\subsection{Unilateral prediction interval} \label{sec_uni_pred_int}

We are first interested in the determination of a unilateral prediction interval.
This allows us to define a limit value that a variable cannot exceed (or reach, 
depending on the context) with a given probability level. 
In the real-life radiological context, this can then be used to estimate, on the 
basis of a small number of contaminant measures, the quantity of contaminant which does not exceed a safety threshold value.

Mathematically, a unilateral prediction interval for a random variable $X \in \mathbb{R}$ is:
\begin{equation} \label{eq_inequality}
\PP(X \geq s) \leq \alpha ,
\end{equation}
where $s \in \mathbb{R}$ is the threshold value and $\alpha \in [0,1]$ is the risk probability.
For an absolutely continuous random variable, this is equivalent to the following inequality:
\begin{equation} \label{eq_inequality2}
\PP(X \leq s) \geq 1 - \alpha = \gamma  \; \; .
\end{equation}
In other words, $s$ is a quantile of $X$ of order greater than $\gamma$.
%which will be used while considering the $\gamma$-quantile of $X$.

In the following sections, we introduce some theoretical inequalities which require the existence (and sometimes the knowledge) of the mean $\mu$ and  standard deviation $\sigma$ of $X$.
Such inequalities are of the following form:
\begin{equation} \label{eq_predint}
\PP(X \geq \mu + t) \leq \left(1 + \frac{t^2}{k\sigma^2}\right)^{-1} ,
\end{equation}
where $t \geq 0$ and $k$ is a positive constant.\\
\\
{\bf General hypothesis of all of these} inequalities: $X$ is absolutely continuous with finite mean and variance.\\
\\
{\bf Hypothesis related to applying them}: The sample from $X$ is i.i.d. (independent and identically distributed).

%%%%%%%%%%%%%%%%%%%%%%%%%%%%%%%%%%%%%%%%%%%%
\subsection{The Gaussian approximation}\label{sec:gauss}

Provided that the random variable $X$ is normally distributed, the derivation of an unilateral prediction interval related to a given risk probability $\alpha$ depends on the knowledge or the ignorance of the mean and standard deviation of $X$.
 
%%%%%%%%%%%%%%%%%%%%
\subsubsection{Known moments}

Denoting by $z_u$ the quantile of level $u$ of the standard Gaussian distribution $\mathcal{N}(0,1)$ (whose value can be easily found in standard normal tables or basic statistical software), we get
\begin{equation}
 \PP\left( \frac{X - \mu}{\sigma}  \, \geq \, z_{1-\alpha}\right) = \alpha \; ,
\end{equation} 
which is equivalent to
\begin{equation}
 \PP\left( X  \, \geq \, \mu + \sigma z_{1-\alpha}  \right) = \alpha \; .
\end{equation} 
This relationship is a special case of Eq. (\ref{eq_predint}) in which the right hand side is no longer an upper bound but the actual risk probability $\alpha$, and where $t=\sigma z_{1-\alpha}$. 
It is easy to show that the parameter $k$ is then equal to $z_{1-\alpha}^2 \alpha/(1-\alpha)$.

%%%%%%%%%%%%%%%%%%%%
\subsubsection{Unknown moments: the $k$-factor method}
\label{sec:kFactor}

The $k$-factor method (also called Owen's method in the literature) develops corrected formulas in order to take into account lack of knowledge of the mean and standard deviation of the variable \cite{owe63,schneu09}. 
It provides tolerance intervals for a normal distribution and in the particular case of a unilateral tolerance interval, one can write :
\begin{equation}
\mathbb{P}\left[\mathbb{P}(X \leq \hat{\mu}_n + k\hat{\sigma}_n)\geq 1-\alpha\right] \geq \beta,
\end{equation}
with \begin{equation}
k = \frac{t_{n-1,\beta,\delta}}{\sqrt{n}}
\end{equation}
and 
\begin{equation}
\delta = z_{1-\alpha} \sqrt{n},
\end{equation}
where $n$ is the sample size, $t_{n-1,\beta,\delta}$ is the $\beta\,$-\,quantile of the non-central $t$-distribution with $n-1$ degrees of freedom and non-centrality parameter $\delta$. $\hat{\mu}_n$ and $\hat{\sigma}_n$ are respectively the empirical mean and standard deviation computed from the sample.
It is quite easy to compute $k$ knowing $\alpha$ and $\beta$.
In our context, the problem is more difficult as we have to solve $k(\alpha, \beta) = \frac{s - \hat{\mu}_n}{\hat{\sigma}_n}$ in order to find the risk probability $\alpha$ associated with the confidence $\beta$.

However, the $k$-factor method is also based on the assumption of normality, and strong care must be taken when applying it to distributions other than Gaussian.
In such situations, the application of a Gaussian approximation may provide non-conservative bounds due to the presence of a small sample, especially in the case of a significantly skewed random variable $X$. 

%%%%%%%%%%%%%%%%%%%%
\subsubsection{Transformation to a Gaussian distribution}

If the original sample of the data does not appear to follow a normal distribution, several transformations of the data can be tried to allow the data to be represented by a normal distribution.
The $k$-factor method can therefore be applied to the normal transformed data in order to obtain either the risk probability, either the prediction or tolerance interval (which has to be back-transformed to obtain the right values).

However, for our purpose which is the risk statistical estimation from small samples, we consider this solution not satisfactory because it might be often not applicable.
First, the Box-Cox family of transformations \cite{boxcox64}, which is a power transformation and includes the logarithmic transformation, requires the fitting by maximum likelihood of the transformation tuning parameter. 
The maximum likelihood process is subject to caution for small data samples.
Second, the iso-probabilistic transformation (see for example \cite{lema09}), also called the Nataf transformation or the Gaussian anamorphosis, consists in applying the data inverse distribution function to the sample.
A such distribution transformation requires the empirical cumulative distribution function, which is built from the data.
With small-size samples, the empirical distribution function is a coarse approximation of the true distribution function and the resulting transformation would be in doubt.

In conclusion, we consider that this solution is inadequate because we cannot guarantee that the transformation to a Gaussian distribution is valid due to the small sample size.
In particular, the normality of the probability distribution tail, which is our zone of interest, would be largely subject to caution.
Moreover, for the same reason, validating the Gaussian distribution of the transformed data seems irrelevant by statistical tests \cite{dagste86}.
We recall that the validation issue of the inequality remaining hypothesis is essential in our context. 
In the following, what are known as concentration inequalities are presented in order to help determine conservative intervals without the Gaussian distribution assumption.

%%%%%%%%%%%%%%%%%%%%%%%%%%%%%%%%%%%%%%%%%%%%%%%%%%%%%%%%%%%%%%%%%%%%%%%%
\subsection{Concentration inequalities}\label{Robust}

In probability theory, concentration inequalities relate the tail probabilities of a random variable to its statistical central moments\footnote{The first moment of a random variable $X$ is the mean $\mu=E(X)$; the second about the mean is the variance $\sigma^2=E[(X-\mu)^2]$; the third  is the skewness coefficient $\gamma_1=E\left[ \left( \frac{X-\mu}{\sigma}\right)^3\right]$; the fourth  is the kurtosis  $\gamma_2=E\left[ \left( \frac{X-\mu}{\sigma}\right)^4\right]$. }.
Therefore, they provide bounds on the deviation of a random variable away from a given value (for example its mean). 
The various inequalities arise from the information we have about the random variable (mean, variance, bounds, positiveness, etc.).
This is a very old research topic in the statistics and probability fields.  For example, \cite{sav61} reviews thirteen such classic inequalities.
New results have been obtained in the previous decades based on numerous mathematical works focused on concentration of measure principles (see for example \cite{boulug13}).
We restrict our work to three classical inequalities that would seem to be most useful for radiological characterization problems with small samples.
Indeed, they only require the mean and variance of the studied variable that does not need to be bounded, with very low assumptions on its probability distribution.

%%%%%%%%%%%%%%%%%%%%
\subsubsection{The Bienaym\'e-Chebychev inequality}
\label{BC}

The Bienaym\'e-Chebychev inequality is written \cite{nel95}:
\begin{equation} 
  \forall \; t \; \geq \; 0 , \quad \PP(X \geq \mu + t) \leq \left(1 + \frac{t^2}{\sigma^2}\right)^{-1} , \label{eq_pred_BT}
\end{equation}
which corresponds to Eq. (\ref{eq_predint}) with $k=1$. 
As $\mu$ and $\sigma$ are unknown in practical situations, they are replaced with their empirical counterparts $\hat{\mu}$ and $\hat{\sigma}$ (i.e., their estimates from the sample values).
This inequality does not require any hypotheses on the probability distribution of $X$.

In fact, Eq. (\ref{eq_pred_BT}) is also known as the Cantelli inequality or Bienaym\'e-Chebychev-Cantelli inequality.
It is an extension of the classical Bienaym\'e-Chebychev inequality (see \cite{boulug13}) where an absolute deviation is considered inside the probability term of Eq. (\ref{eq_pred_BT}).
For the two following inequalities, the same rearrangement is made.\\
\\
{\bf Hypotheses of the Bienaym\'e-Chebychev (BC) inequality: None}

%%%%%%%%%%%%%%%%%%%%
\subsubsection{The Camp-Meidell inequality}
\label{CM}
The Camp-Meidell inequality is given by \cite{mei22,puk94}:
\begin{equation}
  \forall \; t \; \geq \; 0  , \quad \PP(X \geq \mu + t) \leq \left(1 + \frac{9}{4}\frac{t^2}{\sigma^2}\right)^{-1}, \label{eq_pred_CM}
\end{equation}
which corresponds to Eq. (\ref{eq_predint}) with $k=4/9$. 
As $\mu$ and $\sigma$ are unknown in practical situations, they are replaced with their empirical counterparts $\hat{\mu}$ and $\hat{\sigma}$.

It is interesting to note that this inequality, in its two-sided version, justifies the so-called ``three-sigma rule''.
This rule is traditionally used in manufacturing processes, as it states that $95\%$ of a scalar-valued product output $X$ is found in the interval $[\mu-3\sigma,\mu+3\sigma]$.
In fact, it has been shown in \cite{vyspet80} that this rule is only valid for an output $X$ following a unimodal distribution. 
Indeed, one proof of the expression (\ref{eq_pred_CM}) is based on the bounding of the distribution function by a linear one \cite{van51}.
Furthermore, this inequality requires the hypothesis that the probability distribution of $X$ is differentiable, as well as unimodality of the probability density function (pdf) of $X$.
If so, it can then be applied to all of the unimodal continuous probability distributions used in practice (uniform, Gaussian, triangular, log-normal, Weibull, Gumbel, etc.).\\
\\
{\bf Hypothesis of the Camp-Meidell (CM) inequality: Unimodality of the pdf}

%%%%%%%%%%%%%%%%%%%%
\subsubsection{Van Dantzig inequality}
\label{VD}
The Van Dantzig inequality is given by \cite{van51}:
\begin{equation}
  \forall \; t \; \geq \; 0  , \quad \PP(X \geq \mu + t) \leq \left(1 + \frac{8}{3}\frac{t^2}{\sigma^2}\right)^{-1} , \label{eq_pred_VD}
\end{equation}
which corresponds to Eq. (\ref{eq_predint}) with $k=3/8$. 
As $\mu$ and $\sigma$ are unknown in practical situations, they are replaced with their empirical counterparts $\hat{\mu}$ and $\hat{\sigma}$.

We note that this inequality is relatively unknown,
which may be explained by the only minor improvement obtained with respect to the CM inequality.
One proof of the expression (\ref{eq_pred_VD}) is based on  bounding the distribution function by a quadratic function \cite{van51}.
This inequality requires the hypothesis of second-order differentiability of  the probability distribution of $X$, and convexity of the density function of $X$.
In fact, it can be applied to all  unimodal continuous probability distributions in their convex parts. Indeed,
the tail of most classical pdfs is convex, including for example the exponential distribution's density function (convex everywhere), the Gaussian, the log-normal, the Weibull, and so on.
Note however that it is not valid for uniform variables.\\
\\
{\bf Hypothesis of the Van Dantzig (VD) inequality: Convexity of the pdf's tail}

%%%%%%%%%%%%%%%%%%%%%%
\subsubsection{Conservative estimates based on bootstrapping}\label{sec:bootstrap}

Application of the three above concentration inequalities requires knowledge of the mean $\mu$ and  standard deviation $\sigma$ of the variable under consideration.
In most practical situations, these quantities are unknown and are directly estimated from their sample counterparts.
However, low confidence is associated with these estimates when dealing with small sample situations: substituting the actual moments by their sample estimates can in fact lead to 
overly optimistic results. 
To overcome this problem, we propose a penalized approach based on bootstrapping,  a common tool in statistical practice \cite{efrtib93}. 

The principles of  bootstrap variation which are used in this work are as follows: for a given sample of size $n$, we generate a large number $B$ of resamples, i.e., samples made of $n$ values selected randomly with replacement from the original sample. 
We then compute the empirical mean and standard deviation of each resample, and apply the inequality of interest.
We thus obtain $B$ resulting values, and can compute certain statistics such as high quantiles (of order $\beta$, which is the confidence value).
We can take for example the $95\%$-quantile (i.e., $\beta=0.95$) to derive a large and conservative value.

%%%%%%%%%%%%%%%%%%%%%%%%%%%%%%%%%%%%%%%%%%%%%%%%%%%%%%%%%%%%%%%%%%%%%%%%
\subsection{Using Wilks' formula}\label{wilks}

We consider the quantile estimation problem of a random variable as stated in Eq. (\ref{eq_inequality2}), where $\gamma=1-\alpha$ is the order of the quantile. 
This problem is equivalent to the previous one of risk bound estimation (Eq. (\ref{eq_inequality})).
The classic (empirical) estimator is based on order statistic derivations \cite{davnag03} of a Monte-Carlo sample.
With small sample size (typically less than $100$ observations), this estimator gives very imprecise quantile estimates (i.e., with large variance), especially for low (less than $5\%$) and large (more than $95\%$) $\beta$ \cite{cangar08}. 

Another strategy consists of calculating a tolerance limit instead of a quantile, using certain order statistics theorems \cite{davnag03}. 
For an upper bound, this provides a upper limit value of the desired quantile with a given confidence level (for example $95\%$). 
Based on this principle, Wilks' formula \cite{wil41,nutwal04} allows us to precisely determine the required sample size in order to estimate, for a random variable, a quantile of order $\alpha$ with confidence level $\beta$. 
This formula was introduced in the nuclear engineering community by the German nuclear safety institute (GRS) at the beginning of the 1990s \cite{hof90}, and then used for various safety assessment problems (see for example \cite{decbaz08}, \cite{ziodim08} and \cite{perioo08}).

We restrict our explanations below to the one-sided case.
Suppose we have an i.i.d. $n$-sample $X_{1},X_{2},\ldots,X_{n}$ drawn from a random variable $X$. We note $M=\mbox{max}_{i}(X_{i})$.
For $M$ to be an upper bound for at least $100 \times \gamma \%$ of possible values of $X$ with given confidence level $\beta$, we require
\begin{equation}
\PP[\PP(X\leq M)\geq\gamma]\geq\beta \;.
\end{equation}
Wilks' formula implies that the sample size $n$ must therefore satisfy the following inequality:
\begin{equation}\label{eq:Wilksunilateral}
1-\gamma^{n}\geq\beta\;.
\end{equation}
In Table \ref{tab:wilks}, we present several consistent combinations of the 
sample size $n$, the quantile order $\gamma$, and the confidence level $\beta$.
\begin{table*}[!ht]
$$\begin{tabular}{|c||c|c|c|c|c|c|c|c|c|}
\hline
{ $\gamma$} & 0.9 & { 0.9} & { 0.9} & 0.95 & 0.95 & { 0.95} & { 0.95} & { 0.99} & { 0.99}\\
\hline 
{ $\beta$} & 0.5 & { 0.9}  & { 0.95} & 0.4 & 0.5  & { 0.78} & { 0.95} & { 0.95} & { 0.99}\\
\hline 
{ $n$}     & 7   & { 22}   & { 29} & 10    & 14   & 30      & { 59}    & { 299}  & { 459}\\
\hline
\end{tabular}$$
\caption{Examples of values consistent with the first-order case via Wilks' formula  (Eq. (\ref{eq:Wilksunilateral})).}\label{tab:wilks}
\end{table*}
%\begin{table*}[!ht]
%$$\begin{tabular}{|c||c|c|c|c|c||c|c|c|c|c|}
%\cline{2-11} 
%\multicolumn{1}{c||}{} & \multicolumn{5}{c||}{{ Two-sided case, first-order}} & \multicolumn{5}{c|}{{ One-sided case, first-order}}\\
%\hline \hline
%{ $\alpha$} & { 0.9} & { 0.9} & { 0.95} & { 0.99} & { 0.99} & { 0.9} & { 0.9} & { 0.95} & { 0.99} & { 0.99}\\
%\hline 
%{ $\beta$} & { 0.9} & { 0.95} & { 0.95} & { 0.95} & { 0.99} & { 0.9} & { 0.95} & { 0.95} & { 0.95} & { 0.99}\\
%\hline 
%{ $n$} & {{ 38}} & {{ 46}} & {{ 93}} & {{ 473}} & {{ 662}} & { 22} & { 29} & { 59} & { 299} & { 459}\\
%\hline
%\end{tabular}$$
%\caption{Examples of values given in the first-order case by Wilks formula  (equations (\ref{eq:Wilksunilateral}) and (\ref{eq:Wilksbilateral})).}\label{tab:wilks}
%\end{table*}

%
%In the two-sided case, so that at least $\gamma \times 100\%$ of possible values of $X$  are between $m$ and $M$ for given confidence level $\beta$, we require
%\begin{equation}
%\PP[\PP(m\leq X\leq M)\geq\alpha]\geq\beta\;.
%\end{equation}
%In this case, the sample size $n$ must satisfy the following inequality:
%\begin{equation}\label{eq:Wilksbilateral}
%1-\gamma^{n}-n(1-\gamma)\gamma^{n-1}\geq\beta\;.
%\end{equation}

Eq. (\ref{eq:Wilksunilateral}) is a first-order equation because the upper 
bound is set equal to the maximum value of the sample. 
To extend Wilks' formula to higher orders, we consider the $n$-sample of the random variable $X$ sorted into increasing order:  $X_{(1)}$ $\leq$ $X_{(2)}$ $\leq$ \ldots $\leq$ $X_{(r)}$ $\leq$ \ldots $\leq$ $X_{(n)}$ ($r$ is the rank). 
For all $1\leq r \leq n$, we set 
\begin{equation}
G(\gamma)=\PP[\PP( X\leq X_{(r)}) \geq \gamma].
\end{equation}
According to  Wilks' formula, the previous equation can be recast as
\begin{equation}\label{eq:wilks}
G(\gamma)=\sum_{i=0}^{r-1}C_{n}^{i}\gamma^{i}(1-\gamma)^{n-i} \;.
\end{equation}
The value $X_{(r)}$ is an upper-bound of the $\gamma$-quantile with confidence level $\beta$ if $1-G(\gamma)\geq\beta$.
%To get the formula in the one-sided case, it suffices to take $s=0$.

Increasing the order when using Wilks' formula helps reduce the variance in the quantile estimator, the price being the requirement of a larger $n$ (according to formula (\ref{eq:wilks})  with  $\beta=1-G(\gamma)$ and fixed $\gamma$). 
%It is interesting to note the relative sample sizes for the one-sided and two-sided cases: the minimum size given by the  Wilks formula for order $q$ in the two-sided case is the same as the minimum size given for order $2q$ in the one-sided case.
Wilks' formula can be used in two ways:
\begin{itemize}
\item
When the goal is to determine the sample size $n$ to be measured for a given $\gamma-$quantile with a given level of confidence $\beta$, the formula (\ref{eq:Wilksunilateral}) can be used with the fixed order $o$ (corresponding to the $o^{th}$ greatest value, $o=n-r+1$): First order ($o=1$) gives $r=n$ (maximal value for the quantile), second order ($o=2$) gives $r=n-1$ (second largest value for the quantile), etc.
\item
When a sample of size $n$ is already available, then the formula (\ref{eq:Wilksunilateral}) can be used to determine the pairs $(\alpha,\beta)$ and the orders $s$ for the estimation of the Wilks quantile.
\end{itemize}
{\bf Hypotheses of the Wilks formula: None.}

%%%%%%%%%%%%%%%%%%%%%%%%%%%%%%%%%%%%%%%%%%%%%%%%%%%%%%%%%%%%%%%%%%%%%%%%
\section{Numerical tests}

%%%%%%%%%%%%%%%%%%%%
\subsection{Introduction}

The goal of this section is to assess the degree of conservatism of the various approaches presented above, namely the Gaussian-approximation based inequality (Section~\ref{sec:gauss}),  variations of the concentration inequalities (Section~\ref{Robust}) and the Wilks method (Section~\ref{wilks}). To this end, we consider four probability 
distributions which are assumed to generate the data:
\begin{itemize}
\item one Gaussian distribution with mean 210 and standard 
deviation 20,
\item three log-normal distributions with standard deviations equal to 30, 50 
and 70, and with means calculated in such a way that the density maxima
(i.e., the modes) be all equal to 210.
\end{itemize}
We recall that a random variable $X$ is said to be log-normal if $\log(X)$ is normal. 
Note also that the log-normal distribution is classically used for modeling environmental data, such as pollutant concentrations.
Moreover, the hypotheses for all of the concentration inequalities above are valid for these distributions.

\begin{figure}[!ht]
	\centering
	\includegraphics[width=8cm]{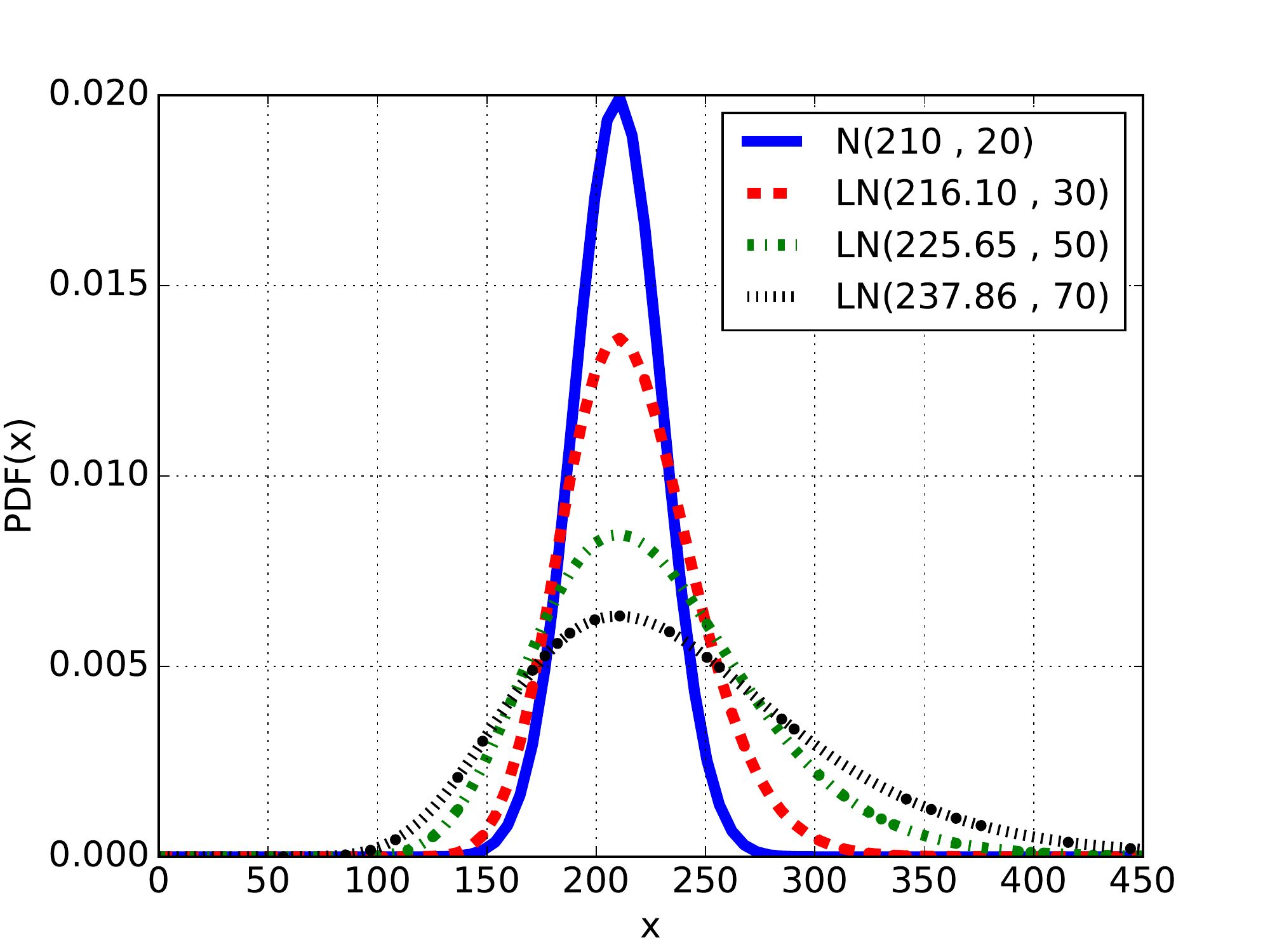} 
	\caption{Four different theoretical pdfs for the random variable $X$.}
	\label{fig:PDFs}
\end{figure}

As shown in Figure~\ref{fig:PDFs}, the distributions exhibit various levels of skewness. The tests carried out for the most skewed distributions 
are able to challenge the robustness of the probabilistic inequalities.

For each theoretical distribution, we want to estimate the minimum probability 
that a value randomly drawn from this same distribution exceeds a given 
threshold $s$. This probability corresponds to the variable $\alpha$ 
in Eq. (\ref{eq_inequality}), and the threshold $s$ to the quantity $\mu+t$ in Eqs. (\ref{eq_pred_BT}), (\ref{eq_pred_CM}) and (\ref{eq_pred_VD}). 
In 
other words, the problem can be cast as follows:
$$\mbox{Estimate~} \alpha \mbox{~such that~} \PP(X \geq s) \leq \alpha ,$$
where $X$ is a random variable following one of the four theoretical 
distributions.

The numerical analysis is organized into two parts:
\begin{enumerate}
\item \emph{The distribution's moments are assumed to be perfectly known.} Thus 
we can compute the exact values of the $\alpha$-estimates given by the concentration 
inequalities (Eqs.(\ref{eq_pred_BT}), (\ref{eq_pred_CM}) and (\ref{eq_pred_VD})). 
We can also estimate $\alpha$ using the Gaussian assumption. However, it does 
not make sense to use Wilks' formula or the $k$-factor method at this stage since the sample 
uncertainty is not taken into account.
\item \emph{The distribution's moments are considered to be unknown (realistic 
case).} The moments are estimated from the data sample at hand. Hence the 
$\alpha$ estimates are affected by  uncertainty in the sample. In other 
words, these estimates are random, and  can be characterized by their 
statistical distributions. In this context, it is relevant to quantify the 
probability that the estimates under-predict or over-predict the $\alpha$ value 
obtained theoretically for each method.
\end{enumerate}

%%%%%%%%%%%%%%%%%%%%
\subsection{Analysis with known moments}

The methods reviewed in this section are the three concentration inequalities and 
the Gaussian approximation. For the concentration inequalities, the risk $\alpha$ is 
estimated as follows:
\begin{equation}
   \alpha \; = \; \left(1+\frac{t^2}{k \sigma^2}\right)^{-1}  \mbox{ with } t \, = \, s - \mu ,
\end{equation}
where $\mu$ and $\sigma$ denote the \emph{exact} values of the distribution's 
mean and standard deviation. For the Gaussian approximation, $\alpha$ is 
estimated through the evaluation of the cumulative density function of the 
Gaussian random variable with mean $\mu$ and standard deviation $\sigma$. The 
estimates are compared to the actual probabilities that any random output value 
exceeds the threshold $s$. A different value for $s$ is given for each 
distribution, so that the probability $\alpha$ of exceeding the threshold is neither too low nor 
too high. $s$ is chosen as the quantile of order $95\%$ of the distribution 
under consideration (accordingly, the actual value of $\alpha$ is $5\%$).

The estimates of the risk $\alpha$ are shown in 
Table~\ref{tab_test_res_known_moments}. We observe that in accordance with the 
theory, the concentration inequalities are always conservative, constantly overestimating the actual risk level. As 
expected, the BC formula is the most conservative, followed 
by the CM one and then the VD one. 
The CM case is particularly interesting because a factor of two is gained over the BC one.
The degree of  conservatism decreases as the distribution skewness increases. The Gaussian 
approximation is of course exact when the distribution is itself Gaussian, 
but it provides overly optimistic estimates of $\alpha$ when the distribution is 
not Gaussian. 

\begin{table*}[!ht]
	\caption{Estimates of the risk $\alpha$ obtained from  
	the Gaussian approximation and the concentration inequalities. The true risk is 
	equal to 5\%.}
	\label{tab_test_res_known_moments}
	\centering
		\begin{tabular}{|c|cccc|}
		\hline
		\multirow{ 2}{*}{~~~~~Methods~~~~~}              & \multicolumn{4}{c|}{Distributions for $X$} \\
		 & $\mathcal{N}(210,20)$ & $\mathcal{LN}(216.10,30)$ & $\mathcal{LN}(225.65,50)$  & $\mathcal{LN}(237.86,70)$  \\ 
    \hline  
	Gauss &  0.05 & 0.04 & 0.04 & 0.03\\
	BC &  0.27 & 0.25 & 0.23 & 0.23 \\
	CM & 0.14 & 0.13 & 0.12 & 0.12 \\
	VD & 0.12 & 0.11 & 0.10 & 0.10 \\
    
	\hline
    \end{tabular}
\end{table*}

%%%%%%%%%%%%%%%%
\subsection{Analysis with unknown moments}

%%%%%%%
\subsubsection{Sample moments-based risk estimates}

In practice, the distribution of the population from which were generated the 
data is unknown, and hence the distribution's mean and standard 
deviation. Therefore, these parameters have to be estimated from the data sample 
at hand in order to derive the risk of exceeding the threshold $s$. This induces 
some randomness in  estimates of the risk $\alpha$ since the sample is 
itself random. As a consequence, it may be possible to greater or less noticeably 
underestimate $\alpha$ in practical situations. 

In order to study the level of conservatism of the various approaches when 
subject to sample randomness, we estimate the statistical distribution of 
the estimates and look at the proportion of non-conservative estimates, i.e., 
estimates less than the actual value $\alpha=5\%$. We only consider the most 
skewed distribution in this section, i.e., the log-normal with mean 
237.86 and standard deviation 70. 
From the theoretical distribution, we randomly 
draw a large number $N$ of samples of a given size $n$. In this study, we choose $N=5,000$ and 
$n \in \{10,30\}$. For a given distribution and a given sample size $n$, $N$ 
risk estimates are computed from the various methods, leading to a sample of 
$N$ estimates of $\alpha$. The empirical distribution of this sample is 
compared to the true risk $5\%$, and the proportion of non-conservative 
estimates out of the $N$ values is computed.

\begin{figure}[!ht]
	\centering
	\includegraphics[width=8cm]{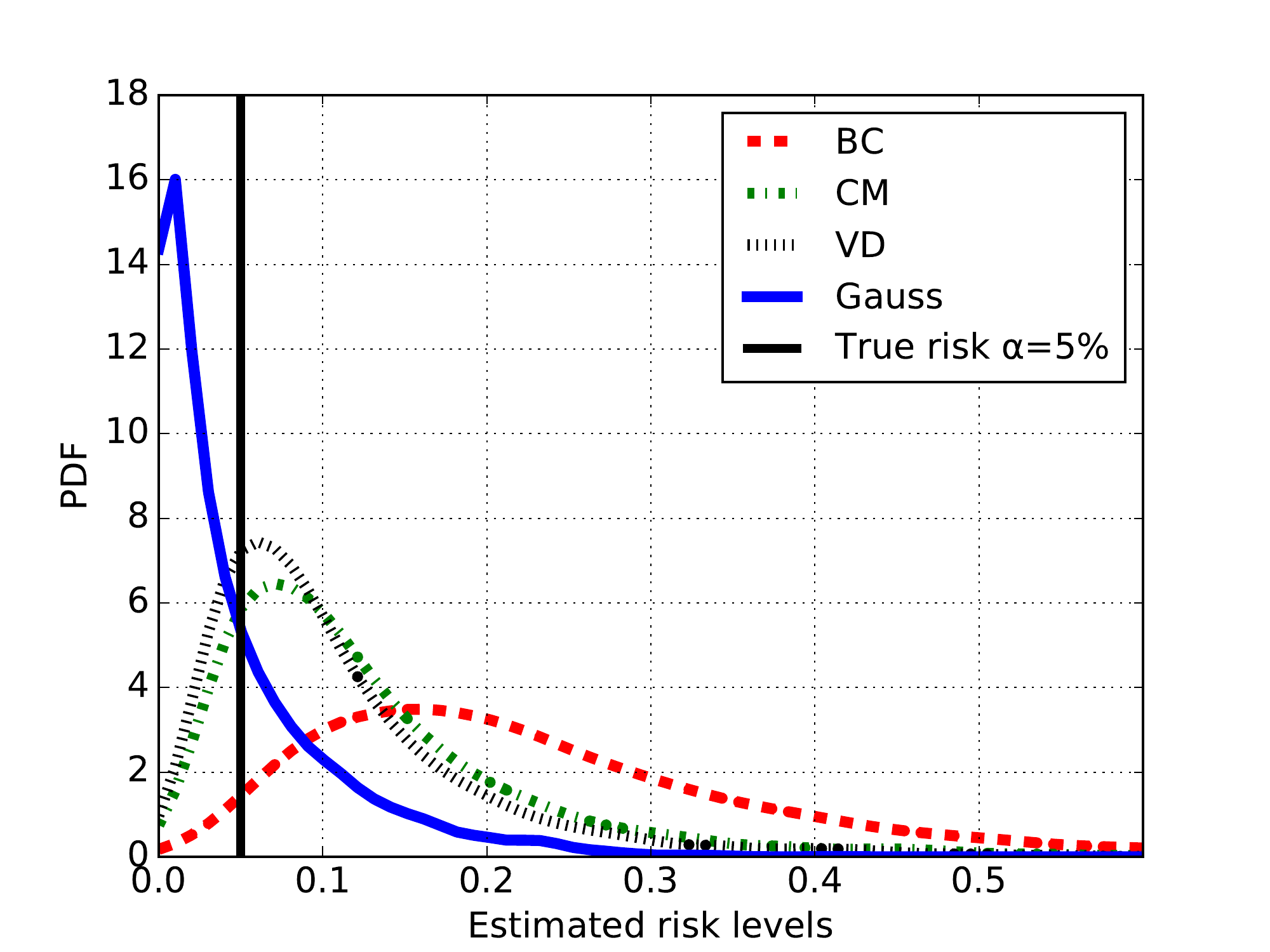} \\
	\vspace{0.2cm}
	\begin{tabular}{|C{1.7cm}|C{1.7cm}|C{1.7cm}|C{1.7cm}|}
	\hline
	\multicolumn{4}{|c|}{Proportion of non-conservative estimates} \\
	\hline
	Gauss & BC & CM & VD \\
	\hline
	0.68 & 0.02 & 0.16 & 0.22 \\
	\hline
	\end{tabular}
	\vspace{0.2cm}
	\caption{Statistical distributions of the $\alpha$ estimates based on 
	samples of size $n=10$, for the log-normal distribution with mean 237.86 and 
	standard deviation 70. The actual risk is equal to 5\%.}
	\label{fig:PDFs:estimates:n10}
\end{figure}

\begin{figure}[!ht]
	\centering
	\includegraphics[width=8cm]{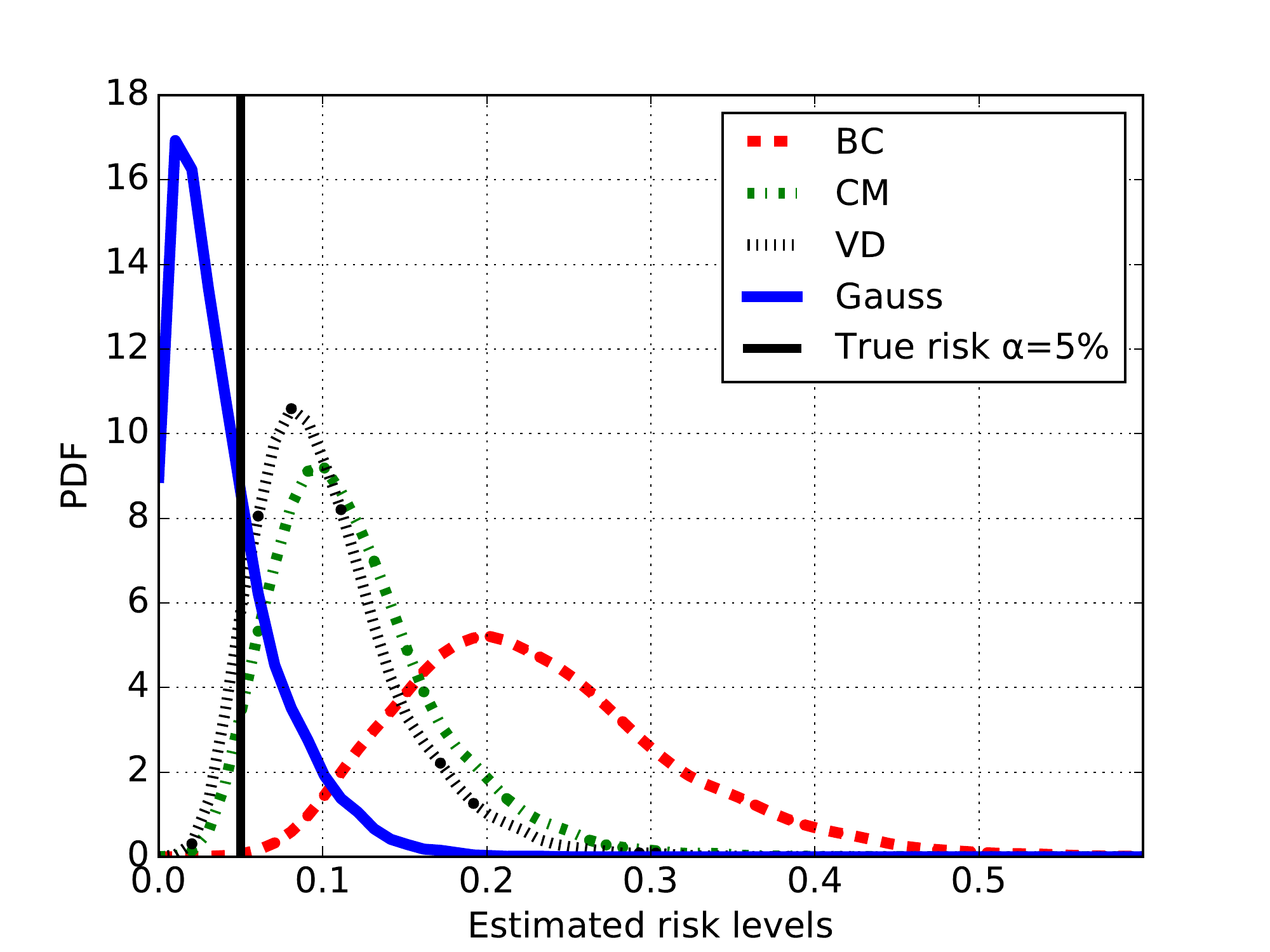}
	\vspace{0.2cm}
	\begin{tabular}{|C{1.7cm}|C{1.7cm}|C{1.7cm}|C{1.7cm}|}
	\hline
	\multicolumn{4}{|c|}{Proportion of non-conservative estimates} \\
	\hline
	Gauss & BC & CM & VD \\
	\hline
	0.71 & 0.00 & 0.03 & 0.06 \\
	\hline
	\end{tabular}
	\vspace{0.2cm} 
	\caption{Statistical distributions of the $\alpha$ estimates based on 
	samples of size $n=30$, for the log-normal distribution with mean 237.86 and 
	standard deviation 70. The actual risk is equal to 5\%.}
	\label{fig:PDFs:estimates:n30}
\end{figure}

The results obtained with $n=10$ and $n=30$ are represented in 
Figures~\ref{fig:PDFs:estimates:n10} and~\ref{fig:PDFs:estimates:n30}, 
respectively. It appears that the Gaussian approximation strongly underestimates 
the actual risk level, with about $70\%$ of its results being smaller than the 
reference risk value $0.05$. (note that the method provided many 
negative values for risk levels that were truncated to zero.) The concentration inequalities turned out to be 
much more conservative, especially the BC one. Nonetheless, 
when dealing with samples of size $n=10$, the CM and VD formulas 
yield a non-negligible proportion of overly optimistic estimates of $\alpha$. The 
three types of inequality are more conservative when more data are available, i.e., when the sample size $n$ is set to 30.

These results are compared to the first-order Wilks approach. 
The first-order approach (which means that the threshold is the sample maximum) is taken because the sample sizes are extremely small in these tests.
According to formula~(\ref{eq:Wilksunilateral}), for an actual risk of 
level $\alpha=0.05$ and a sample size $n=10$, the proportion of 
non-conservative estimates is less than 0.60 (this value corresponds to the 
quantity $1 - \beta$). This value decreases to 0.21 when the sample size is 
 $n=30$. Thus the Wilks method lies between the Gaussian approximation 
and the concentration inequalities for this test case, in terms of conservatism. 
However, an advantage of the method is that it directly gives an upper bound of
$1 - \beta$ of the risk of being non-conservative, in contrast to the other 
strategies.  

%%%%%%%%%%%%%%%%%%%%%%
\subsubsection{Penalized risk estimates based on bootstrapping}

We have shown that applying the Gaussian and robust methods by simply 
substituting the actual moments by their sample estimates can lead to 
overly optimistic results. To overcome this problem, we propose a penalized 
approach based on bootstrapping (see Section~\ref{sec:bootstrap}) that we will compare to the $k$-factor method i.e. the Gaussian approximation to get tolerance intervals (see Section~\ref{sec:kFactor}). 
The principles is as follows. For a given sample of size $n$, we generate a large number $B$ of resamples (say, $B = 500$).
We compute the empirical mean and standard deviation of each resample, 
and then derive in each case an estimate of $\alpha$, as shown in the previous section. This results in a bootstrap set of $B$ estimates of $\alpha$. In a 
conservative way, we can compute a high quantile of a given order of this set, say  
5\%. The calculated value serves as an estimate of the risk $\alpha$.

\begin{figure}[!ht]
	\centering
	\includegraphics[width=8cm]{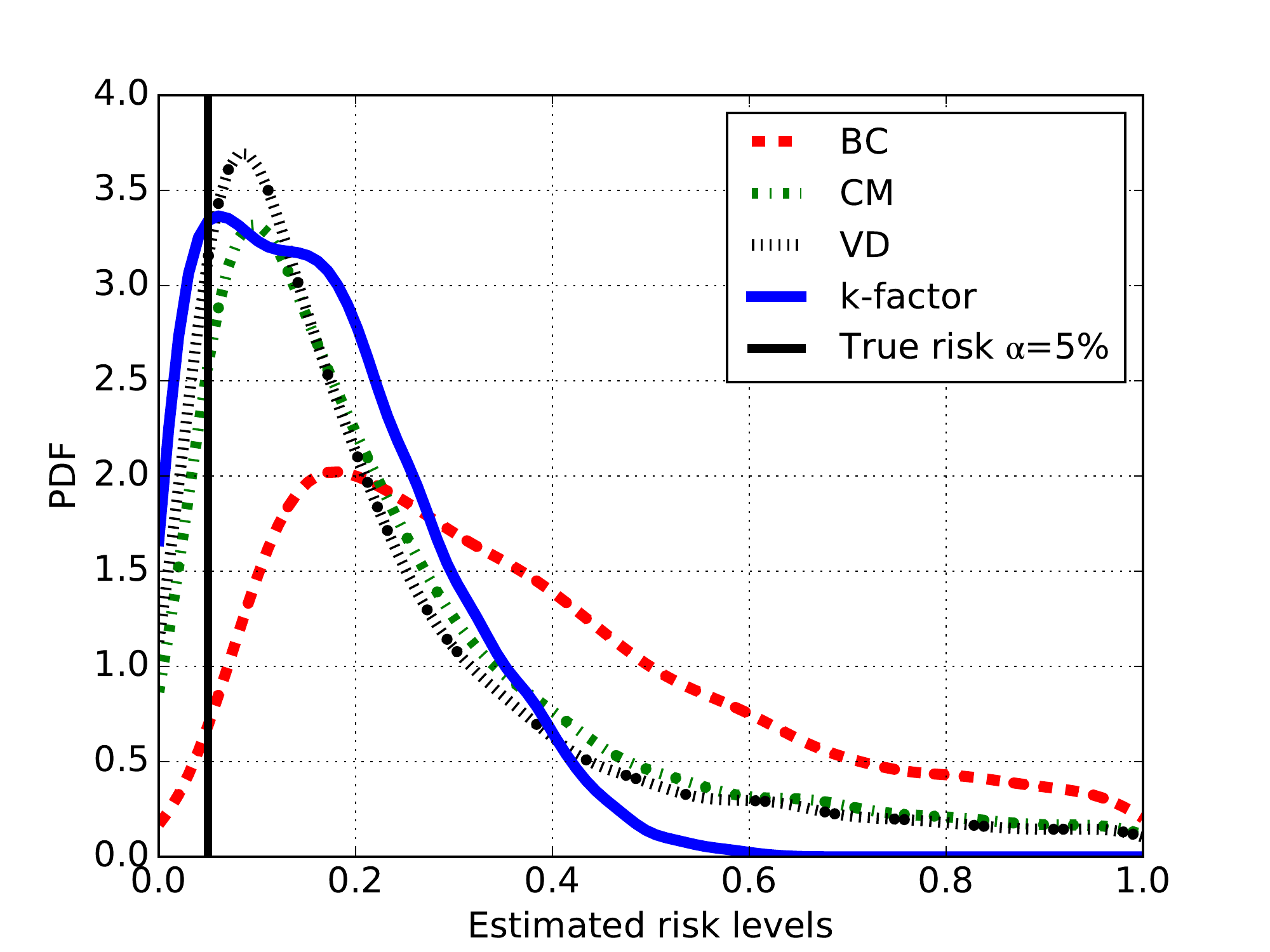} \\
	\vspace{0.2cm}
	\begin{tabular}{|C{1.7cm}|C{1.7cm}|C{1.7cm}|C{1.7cm}|}
	\hline
	\multicolumn{4}{|c|}{Proportion of non-conservative estimates} \\
	\hline
	$k$-factor & BC & CM & VD \\
	\hline
	0.17 & 0.01 & 0.07 & 0.10 \\
	\hline
	\end{tabular}
	\vspace{0.2cm}
	\caption{Statistical distributions of the bootstrap $\alpha$ estimates based on 
	samples of size $n=10$, for the log-normal distribution with mean 237.86 and 
	standard deviation 70. The actual risk is equal to 5\%.}
	\label{fig:PDFs:estimates:n10:boot}
\end{figure}

\begin{figure}[!ht]
	\centering
	\includegraphics[width=8cm]{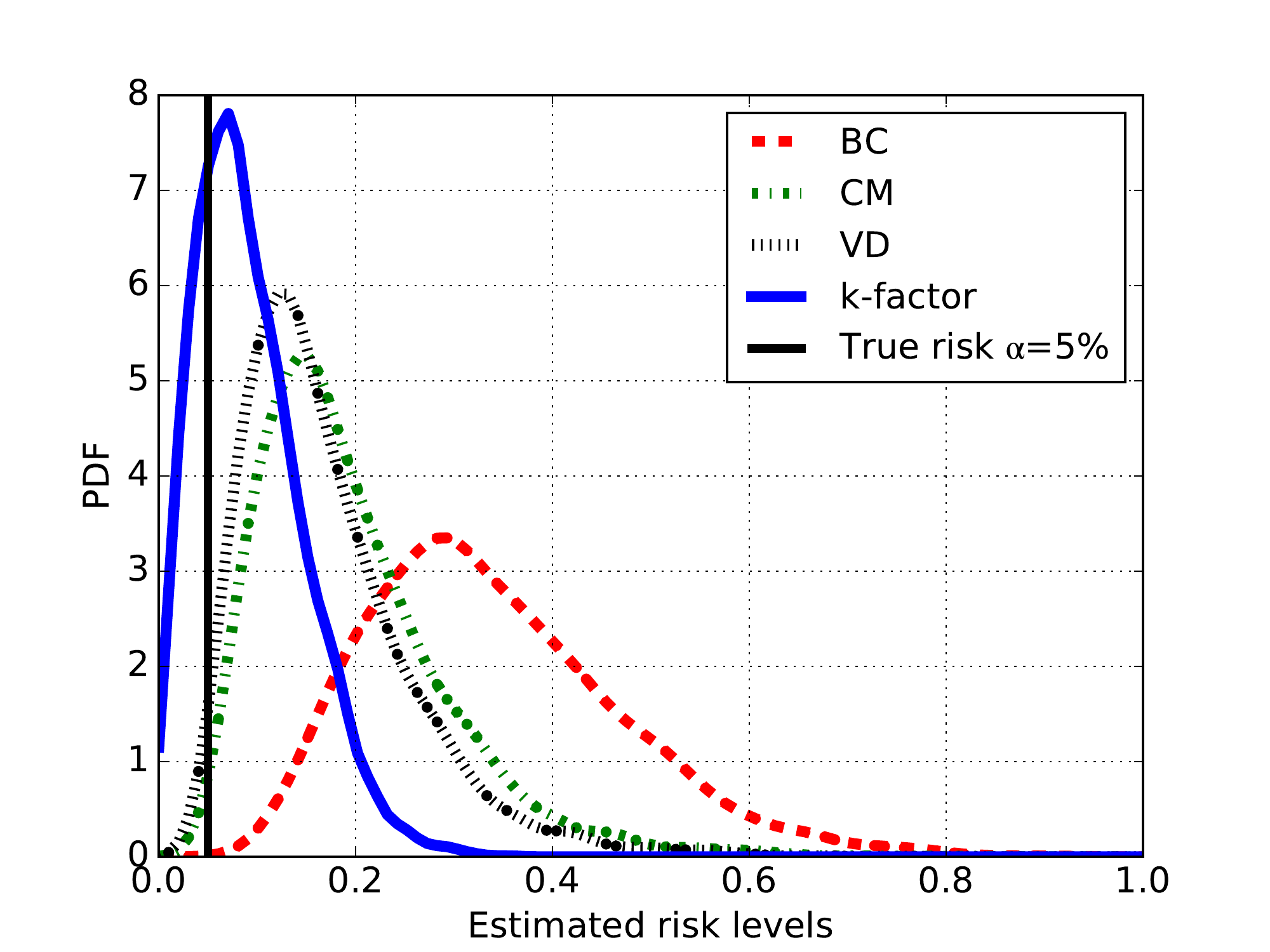}
	\vspace{0.2cm}
	\begin{tabular}{|C{1.7cm}|C{1.7cm}|C{1.7cm}|C{1.7cm}|}
	\hline
	\multicolumn{4}{|c|}{Proportion of non-conservative estimates} \\
	\hline
	$k$-factor & BC & CM & VD \\
	\hline
	0.25 & 0.00 & 0.00 & 0.01 \\
	\hline
	\end{tabular}
	\vspace{0.2cm} 
	\caption{Statistical distributions of the bootstrap $\alpha$ estimates based on 
	samples of size $n=30$, for the log-normal distribution with mean 237.86 and 
	standard deviation 70. The actual risk is equal to 5\%.}
	\label{fig:PDFs:estimates:n30:boot}
\end{figure}

As in the previous section, we focus on the log-normal distribution 
with the greatest skewness. The results related to sample sizes $n$ equal to 10 and 30 are plotted in 
Figures~\ref{fig:PDFs:estimates:n10:boot} and~\ref{fig:PDFs:estimates:n30:boot}, 
respectively. As expected, we see that the bootstrap-based estimators 
are significantly more conservative than their ``moment-based'' counterparts. 
In particular, for $n=10$, the level of conservatism is roughly doubled. For 
$n=30$, all of the concentration inequalities led to very small proportions 
(less than 1\%) of under-prediction of the actual risk $\alpha$. The Gaussian 
approximation however still remains unreliable for the skewed distribution 
under consideration.

The drawback of the conservatism of the robust approaches is that they 
can yield grossly-exaggerated estimations of the risk of exceeding the 
threshold value, especially the BC formula. On the other 
hand, the VD formula relies upon assumptions about the density function 
of the population, which are not easy to check in practice. In the absence of 
further investigation,  combination of the CM inequality with 
a bootstrap penalization may be a reasonable approach.

In conclusion, for all these tests based on small data samples, the inadequacy of the Gaussian approximation has been shown, while concentration inequalities, used in a conservative manner (using a boostrap technique), show strong robustness.
Wilks' formula offers the advantage of directly giving an upper bound of the risk of being non-conservative, but is not as advantageous when dealing with very small sample sizes (low conservatism).
If their assumptions can be considered reasonable, the Camp-Meidell and Van Dantzig inequalities should be used preferentially, as the Bienaym\'e-Chebychev inequality usually gives 
overly-conservative results.
In the following section, we illustrate the practical usefulness of all these tools in real situations.

%%%%%%%%%%%%%%%%%%%%%%%%%%%%%%%%%%%%%%%%%%%%%%%%%%%%%%%%%%%%%%%%%%%%%%%%

\section{Applications}
\label{Appli}

%%%%%%%%%%
\subsection{Case 1: Contamination characterization}

This case study concerns the radiological activity (denoted $X$) of Cesium $137$ in a large-sized population of waste objects.
This characterization enables us to put each waste object in a suitable waste category, e.g., low-activity or high-activity. The reliability of this classification is all the more crucial as it directly affects the total cost of waste management. Indeed, putting objects in the high-activity waste category is much more expensive than in the low-activity one.

\begin{figure*}[!ht]
	\centering
	   \includegraphics[width=16cm,height=7cm]{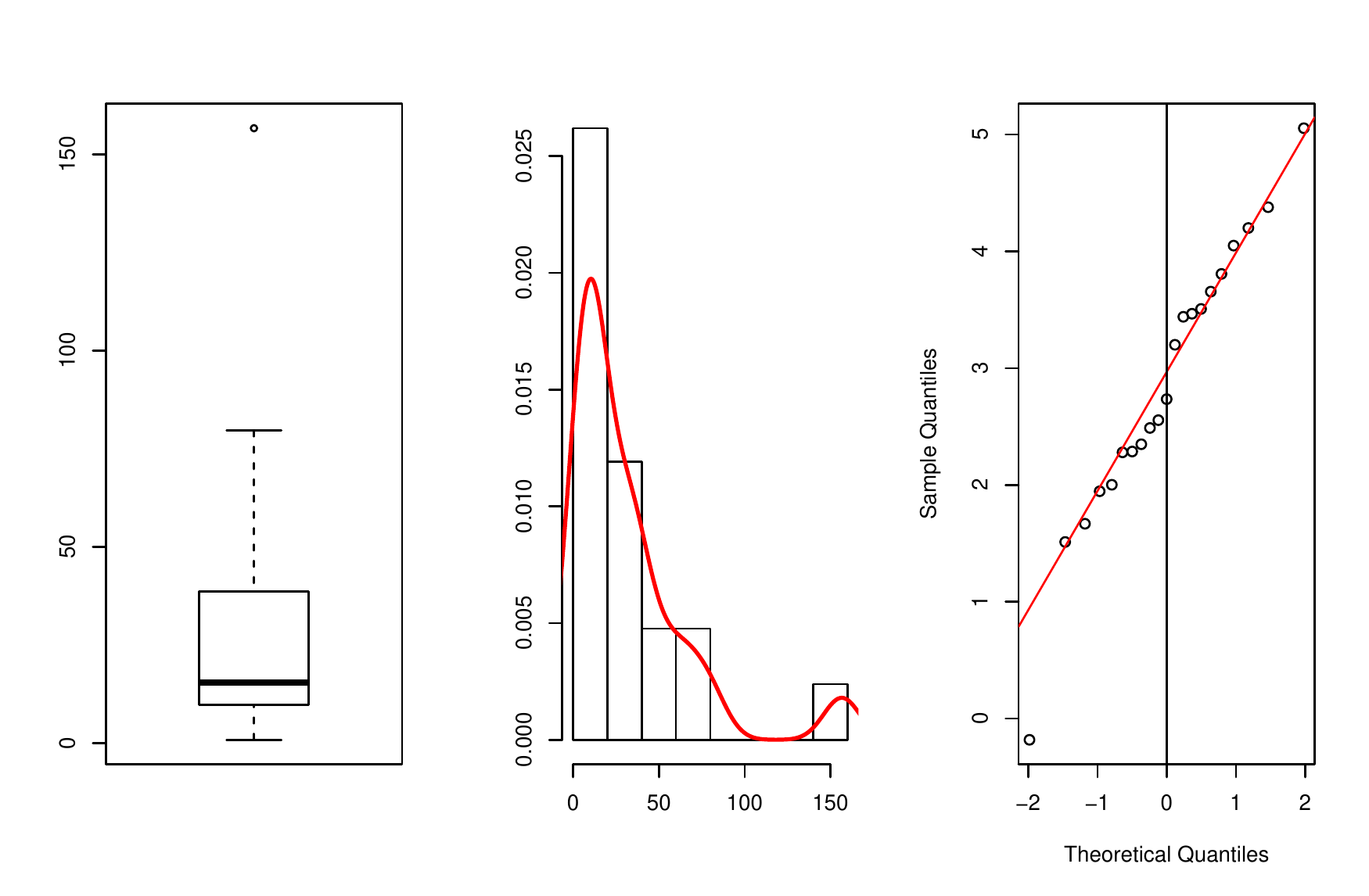} 
	\caption{Case 1 ($21$ $^{137}Cs$ activity measures): Boxplot (left), histogram with a smoothed-kernel density function (middle) and quantile-quantile plot with respect to a log-normal distribution.}
	\label{fig:boxplot-PDF-Cs}
\end{figure*}

A complete characterization of the population of waste objects is impossible, and only $21$ measurements were in fact made.
Reasoning in terms of statistics, it is assumed that this small-sized sample ($n=21$) has been randomly chosen from an unknown infinite population associated 
with some probability distribution.
Each object of this sample has been characterized by its $^{137}Cs$ activity measure (in Bq/cm$^2$).
The summary statistics which are estimated with these data are the following: mean $\hat{\mu}~=~31.45$, median~=~$15.4$, standard deviation $\hat{\sigma}~=~36.11$, Min = $0.83$, Max = $156.67$.
Figure \ref{fig:boxplot-PDF-Cs} shows the boxplot, histogram and smoothed-kernel density of these data.
The distribution resembles a log-normal one, with high asymmetry, a mean much larger than the median, a standard deviation larger than the mean, many low values and a few high ones.
The better quantile-quantile plot is the one obtained with respect to the log-normal distribution (see Figure \ref{fig:boxplot-PDF-Cs}) and then supports this intuition.
However, the size of the sample is too small to be confident on the results of statistical tests which would confirm this \cite{dagste86}.
The extreme value at $156.67$ seems to be isolated from the rest of the sample values, but we have no argument allowing us to consider it as an outlier.
Moreover, the actual data density is considered as unimodal because there is no physical reason to believe that this high value comes from a second population with a different contamination type.
Even if it may be subject to discussion, the hypothesis of convexity of the density's tail can be supposed.

From $21$ activity measures, we want to estimate the proportion of the total population which has a radiological activity larger than a given threshold.
First, the quantity of waste objects whose activity does exceed the threshold $s=100$ Bq/cm$^2$ has to be determined.
This could be an important issue in terms of predefining the volume of this waste category.

We were unsuccessful in fitting (with a high degree of confidence) a parametric statistical distribution (even a log-normal one) to these data.
Thus, only distribution-free tools, as those discussed in this paper, can be used to build prediction intervals with a sufficient degree of confidence.
The probabilistic inequalities of type (\ref{eq_predint}) were then applied, by replacing $\mu$ and $\sigma$ by their estimates:
\begin{equation} \label{eq_predint-est}
\PP(X \geq \hat{\mu} + t) \leq \left(1 + \frac{t^2}{k\hat{\sigma}^2}\right)^{-1} ,
\end{equation}
where $t = s - \hat{\mu}$ and the values of $k$ depend on the inequality ($k=1$ for BC, $k=4/9$ for CM, and $k=3/8$ for VD).
This equation can also be expressed by using $s$:
\begin{equation} \label{eq_predint-est2} 
\PP(X \geq s) \leq \left(1 + \frac{(s-\hat{\mu})^2}{k\hat{\sigma}^2}\right)^{-1} .
\end{equation}

The first row of Table \ref{tab:case1} gives the risk bound results for the various inequalities for the threshold set equal to $100$ Bq/cm$^2$, and using the empirical estimates of $\mu$ and $\sigma$.
The second row provides bootstrap-based conservative estimates of the risk bound by taking the $95\%$ quantile of $B=10,000$ risk bounds estimated from $B$ replicas of the data sample.
The interpretation of these two results reveals that:
\begin{itemize}
\item By the BC inequality, we obtain from (\ref{eq_predint-est2}) that \\
$ \left(1 + \frac{(s-\hat{\mu})^2}{\hat{\sigma}^2}\right)^{-1}=0.2172$, so we coarsely estimate that less than $21.7$\% of the population has an activity larger than $100$ Bq/cm$^2$, and we can guarantee (at a $95\%$ confidence level) that less than $44.8$\% of the population has an activity larger than $100$ Bq/cm$^2$.
\item By the CM inequality, we obtain from (\ref{eq_predint-est2}) that \\
$ \left(1 + \frac{(s-\hat{\mu})^2}{\frac{4}{9}\hat{\sigma}^2}\right)^{-1}=0.1098$, so we coarsely estimate that less than $11$\% of the population has an activity larger than $100$ Bq/cm$^2$, and we can guarantee (at a $95\%$ confidence level) that less than $26.5$\% of the population has an activity larger than $100$ Bq/cm$^2$.
\item By the VD inequality, we obtain from (\ref{eq_predint-est2}) that \\
$ \left(1 + \frac{(s-\hat{\mu})^2}{\frac{3}{8}\hat{\sigma}^2}\right)^{-1}=0.0942$, so we coarsely estimate that less than $9.4$\% of the population has an activity larger than $100$ Bq/cm$^2$, and we can guarantee (at a $95\%$ confidence level) that less than $23.3$\% of the population has an activity larger than $100$ Bq/cm$^2$.
\end{itemize}
This simple application illustrates the gain we can obtain by using the CM or VD inequalities instead of the BC one. Knowing from the bootstrap's conservative estimates that $24\%$ instead of $45\%$ of the waste objects can be classified in the high-activity waste category would help to avoid an 
overly-conservative estimate of the waste management cost.

\begin{table}[!ht]
	\caption{Case 1: Estimates of the risk $\alpha$ obtained from the concentration inequalities and Wilks' formula, for different threshold values $s$.}
	\label{tab:case1}
	\centering
		\begin{tabular}{|c||c|c|c|c|}
		\hline
		$s$ & BC & CM & VD & Wilks\\
    \hline  
    100 & 0.217 & 0.110 & 0.094 & -\\
		\hline
		100 &&&&\\
		$\beta=0.95$ & 0.448 & 0.265 & 0.233 & -\\
		\hline
		156.67 &&&&\\
		$\beta=0.95$ & 0.169 & 0.083 & 0.071 & 0.133\\
		$\beta=0.78$ & 0.118 & 0.056 & 0.048 & 0.071\\
		\hline
		79.67 &&&&\\
		$\beta=0.95$ & 0.665 & 0.469 & 0.427 & 0.207	\\
		$\beta=0.999$ & 0.877 & 0.760 & 0.728 & 0.427\\
		\hline
    \end{tabular}
\end{table} 

Next, we use Wilks' formula to illustrate what kind of statistical information can be inferred for the given data sample. 
For the sample size $n=21$, we can estimate two types of quantile:
\begin{itemize}
\item
A unilateral first-order $\gamma-$quantile with a confidence level $\beta$, and then we deduce $\alpha=1-\gamma$ and $\beta$ from the equation (\ref{eq:Wilksunilateral}).
We obtain the following solutions: \\
$\PP[\PP(X\leq 156.67)\geq 0.896]\geq 0.9 ;\; (\alpha,\beta)=(10.4\%,90\%)$,\\
$\PP[\PP(X\leq 156.67)\geq 0.867]\geq 0.95 ;\; (\alpha,\beta)=(13.3\%,95\%)$.
\\
\item
A unilateral second-order $\gamma-$quantile with a confidence level $\beta$, and then we deduce $\alpha=1-\gamma$ and $\beta$ from the equation (\ref{eq:wilks}) with $o=2$ and $r=n-1$.
We obtain the following potential solutions: \\
$\PP[\PP(X\leq 79.67)\geq 0.827]\geq 0.9  ;\; (\alpha,\beta)=(17.3\%,90\%)$,\\
$\PP[\PP(X\leq 79.67)\geq 0.793]\geq 0.95 ;\; (\alpha,\beta)=(20.7\%,95\%)$.
\end{itemize}

In Table \ref{tab:case1} (rows 3 and 4), we compare these results with those of the concentration inequalities by adjusting the corresponding thresholds $s$.
Indeed, comparisons cannot be made with $s=100$ because Wilks' formula can only be applied with a quantile value coming from the data sample values.
This is the major drawback of  Wilks' formula.
For the low-quantile case ($79.67$, in row 4), i.e., large risk bounds, Wilks' formula  relevance is clear because it  always provides less penalized results than the BC, CM and VD inequalities; the gain is a factor of two.
However, in the high-quantile case ($156.67$, in row 3), i.e., small risk bounds, the CM and VD inequalities provide less penalized results, with $8.3\%$ and $7.1\%$ of the population that may have an activity higher than $156.67$ Bq/cm$^2$ and $13.3\%$ for Wilks' method with a confidence $\beta$ of $95\%$ (resp. $5.6\%$ and $4.8\%$ for CM and VD, $7.1\%$ for Wilks' method with a confidence $\beta$ of $78\%$).
The gain with VD is a factor of two.
The same results are also obtained by giving the values of $\beta$ obtained via Wilks' formula using the conservative $\alpha$ result obtained by the VD inequality.

%The values of $\alpha$ and $\beta$ for a unilateral first order Wilks quantile of a sample of size 21 (which corresponds to the maximal value of the sample) are $(\alpha,\beta)=(89\%,91\%)$ and $(\alpha,\beta)=(90.58\%,87\%)$. The values of $\alpha$ and $\beta$ for a unilateral second order Wilks quantile of a sample of size 21 (which corresponds to the second maximal value of the sample) are $(\alpha,\beta)=(89\%,70\%)$ and $(\alpha,\beta)=(90.58\%,60\%)$.

%%%%%%%%%%%%
\subsection{Case 2: H$_2$ flow rate characterization for drums of radioactive waste}

Some categories of radioactive waste drums may produce hydrogen gas because of the radiolyse reaction of organic matter like PVC, Polyethylene or cellulose mixed with $\alpha$-emitters in the waste. The evaluation of the hydrogen flow rate (denoted $X$ in l/drum/year) produced by radioactive waste drums is required for their disposal in final waste repositories. However, considering the time required for the H$_2$ flow rate measurement of only one drum (more than one month) and the need to characterize a population of several thousand  drums, only a small ($n=38$) randomly chosen sample has been measured.

\begin{figure*}[!ht]
	\centering
	   \includegraphics[width=16cm,height=7cm]{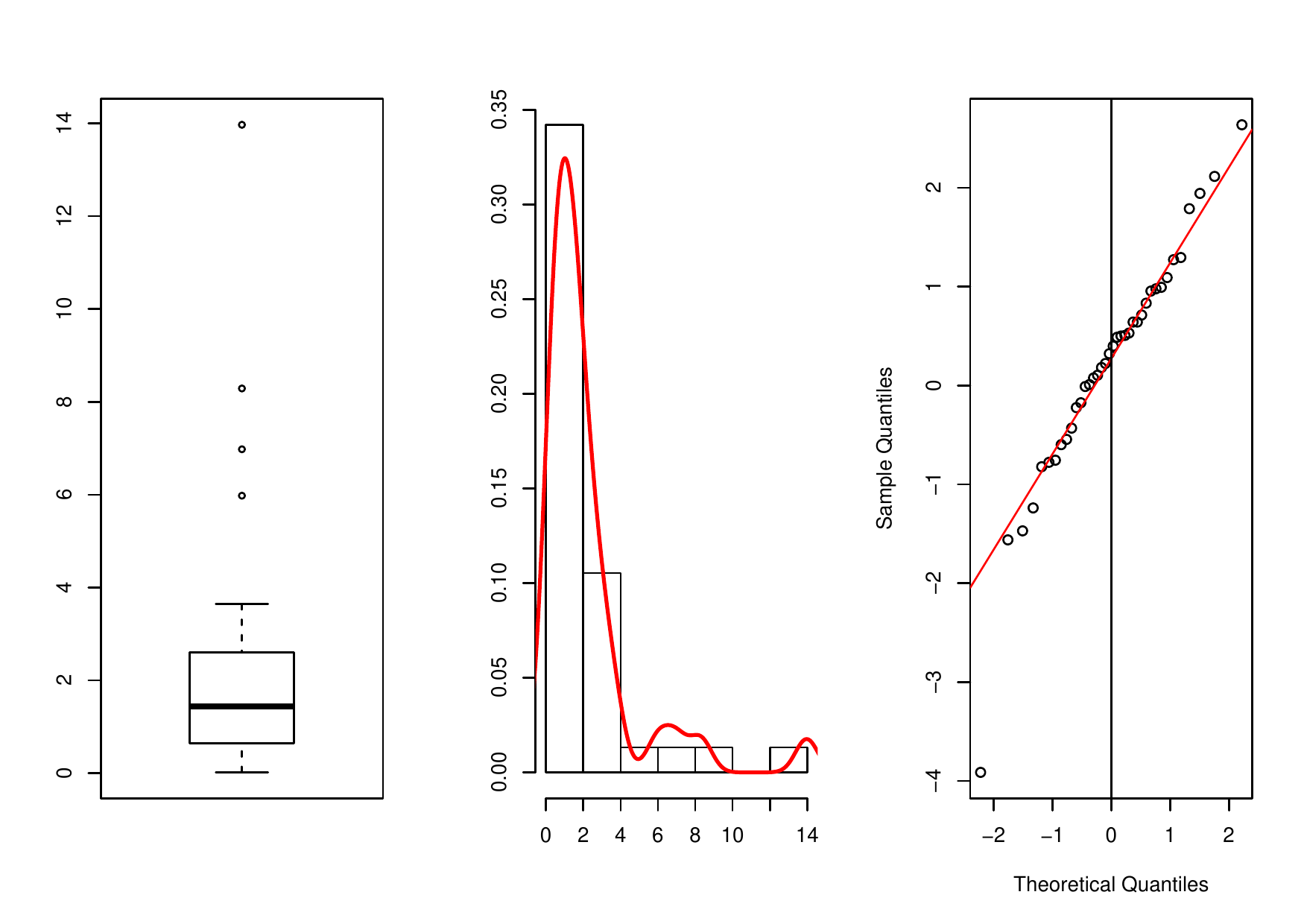} 
	\caption{Case 2 ($38$ hydrogen flow rates): Boxplot (left), histogram with a smoothed-kernel density function (middle) and quantile-quantile plot with respect to a log-normal distribution.}
	\label{fig:boxplot-H2}
\end{figure*}

The summary statistics estimated with the  H$_2$ flow rate data are the following: mean $\hat{\mu}~=~2.18$, median~=~$1.43$, standard deviation $\hat{\sigma}~=~2.67$, Min = $0.02$, Max = $13.97$.
Figure \ref{fig:boxplot-H2} shows the boxplot,  histogram and  smoothed-kernel density of these data. 
As for  Case 1, the distribution looks like a log-normal one, with high asymmetry, a mean much larger than the median, a standard deviation larger than the mean, a lot of low values and a few high ones.
The better quantile-quantile plot is the one obtained with respect to the log-normal distribution (see Figure \ref{fig:boxplot-H2}) and then supports this intuition.
The extreme value at $13.97$ seems to be isolated from the rest of the sample values, but we have no argument that justifies considering it as an outlier.
It tends the distribution to appear as an heavier tail distribution than the log-normal one.
We can directly estimate the 95\%-quantile from the log-normal theoretical distribution that has been fitted $X~=~\mathcal{LN}(0.23,1.16)$:
$$ q_{95\%}=  8.4827 \mbox{ l/drum/year} .$$
However, due to the small number of data that served to fit the pdf, little confidence can be accorded to this value, and justifying it to safety authorities could be difficult.
Moreover, the log-normal distribution is rejected by the Shapiro-Wilks adequacy test (the most robust test for small sample size) with the threshold $5\%$. 
In any case, we are confident that the density can be considered as unimodal and the hypothesis of convexity of the density's tail could also be accepted. 

The first row of Table \ref{tab:case2} gives the risk bound results of the different inequalities for the threshold $s=10$ l/drum/year,  using the empirical estimates of $\mu$ and $\sigma$.
The second row provides bootstrap-based conservative estimates of the risk bound by taking the $95\%$ quantile of $B=10,000$ risk bounds estimated from $B$ replicas of the data sample.
The interpretation of these two results reveals that:
\begin{itemize}
\item By the BC inequality, we coarsely estimate that less than $10.5$\% of the population has an activity larger than $10$ l/drum/year, and we can guarantee (at a $95\%$ confidence level) that less than $21.2$\% of the population has a $H_2$ flow rate larger than $10$ l/drum/year.\\

\item By the CM inequality, we coarsely estimate that less than $5$\% of the population has an activity larger than $10$ l/drum/year, and we can guarantee (at a $95\%$ confidence level) that less than $10.7$\% of the population has a $H_2$ flow rate larger than $10$ l/drum/year.\\

\item By the VD inequality, we coarsely estimate that less than $4.2$\% of the population has an activity larger than $10$ l/drum/year, and we can guarantee (at a $95\%$ confidence level) that less than $9.2$\% of the population has a $H_2$ flow rate larger than $10$ l/drum/year.
\end{itemize}
As for Case 1, the gain we can obtain (here, about $12$\%) by using the CM or VD inequality instead of the BC one is relatively large, in terms of estimating the waste management cost.

\begin{table}[!ht]
	\caption{Case 2: Estimates of the risk $\alpha$ obtained from the concentration inequalities and Wilks' formula, for different threshold values $s$.}
	\label{tab:case2}
	\centering
		\begin{tabular}{|c||cccc|}
		\hline
		$s$ & BC & CM & VD & Wilks\\
    \hline  
    10 & 0.105 & 0.050 & 0.042 & -\\
		\hline
		10 &&&& \\
		$\beta=0.95$ & 0.212 & 0.107 & 0.092 & -\\
		\hline
		13.97 &&&& \\
		$\beta=0.95$ & 0.099 & 0.047 & 0.040 & 0.076\\
		$\beta=0.78$ & 0.070 & 0.032 & 0.027 & 0.040\\
		\hline
		8.29 &&&& \\
		$\beta=0.95$ & 0.315 & 0.170 & 0.147 & 0.119\\
		$\beta=0.98$ & 0.360 & 0.200 & 0.174 & 0.147\\
	\hline
    \end{tabular}
\end{table}

We now use Wilks' formula to illustrate, for the given data sample, what kind of statistical information can be inferred. 
For the sample size $n=38$, we can estimate two types of quantile:
\begin{itemize}
\item
A unilateral first-order $\gamma-$quantile with a confidence level $\beta$, and then we  deduce $\alpha=1-\gamma$ and $\beta$ from the equation (\ref{eq:Wilksunilateral}).
We obtain the following solutions: \\
$\PP[\PP(X\leq 13.97)\geq 0.941]\geq 0.9 ;\; (\alpha,\beta)=(5.9\%,90\%)$,\\
$\PP[\PP(X\leq 13.97)\geq 0.924]\geq 0.95 ;\; (\alpha,\beta)=(7.6\%,95\%)$.
\\
\item
A unilateral second-order $\gamma-$quantile with a confidence level $\beta$, and then we deduce $\alpha=1-\gamma$ and $\beta$ from the equation (\ref{eq:wilks}) with $o=2$ and $r=n-1$.
We obtain the following potential solutions: \\
$\PP[\PP(X\leq 8.29)\geq 0.901]\geq 0.9  ;\; (\alpha,\beta)=(9.9\%,90\%)$,\\
$\PP[\PP(X\leq 8.29)\geq 0.881]\geq 0.95 ;\; (\alpha,\beta)=(11.9\%,95\%)$.
\end{itemize}

In Table \ref{tab:case2} (rows 3 and 4), we compare these results with those of the concentration inequalities, by adjusting the corresponding thresholds $s$.
Indeed, comparisons cannot be made with $s=10$ because Wilks' formula can only be applied with a quantile value coming from the data sample values.
Again, this is the major drawback of  Wilks' formula.
For the low-quantile case ($8.29$, in row 4), i.e., large risk bounds, Wilks' formula is clearly relevant because it  always gives less penalized results than the BC, CM and VD inequalities.
However, in the high-quantile case ($13.97$, in row 3), i.e., small risk bounds, the CM and VD inequalities provide less penalized results; the gain with VD is a factor of two.
The same results are also obtained with the values of $\beta$ obtained from Wilks' formula using the conservative $\alpha$ result obtained by the VD inequality.

%For this case study, the sample size is $n=38$, we can use the Wilks formulas to estimate two types of quantile :
%\begin{itemize}
%\item
%An unilateral first order $\alpha-$quantile with a level of confidence $\beta$, and then we have to deduce $\alpha$ and $\beta$ from the equation (\ref{eq:Wilksunilateral}) and we obtain the following potential solutions: \\
%for $(\alpha,\beta)=(0.92\%,95\%)$, $\PP[\PP(X\leq 13.97)\geq 0.92]\geq 0.95 $,\\
%for $(\alpha,\beta)=(94\%,90\%)$, $\PP[\PP(X\leq 13.97)\geq 0.94]\geq 0.90 $.
%
%\item
%An unilateral second order $\alpha-$quantile with a level of confidence $\beta$, and then we have to deduce $\alpha$ and $\beta$ from the equation (\ref{eq:Wilksunilateral}) with $t=2$ and we obtain the following potential solutions: 
%for $(\alpha,\beta)=(90\%,90\%)$, $\PP[\PP(X\leq 8.3)\geq 0.9]\geq 0.9 $ \\
%for $(\alpha,\beta)=(0.88\%,0.95\%)$, $\PP[\PP(X\leq 8.3)\geq 0.88]\geq 0.95 $.
%
%\end{itemize}
%
%To compare with the results from the concentration inequalities, the values of $\alpha$ and $\beta$ for a unilateral first order Wilks quantile of a sample of size 38 (which corresponds to the maximal value of the sample) are $(\alpha,\beta)=(89.57\%,98\%)$ and $(\alpha,\beta)=(95.8\%,80\%)$. The values of $\alpha$ and $\beta$ for a unilateral second order Wilks quantile of a sample of size 38 (which corresponds to the second maximal value of the sample) are $(\alpha,\beta)=(89.57\%,92\%)$ and $(\alpha,\beta)=(95.8\%,48\%)$.
 
%%%%%%%%%%%%%%%%%%%%%%%%%%%%%%%%%%%%%%%%%%%%%%%%%%%%%%%%%%%%%%%%%%%%%%%%
\section{Conclusion}

As explained in the introduction, a realistic assessment of  risk is of major importance in improving the management of  risk, as well as public acceptance. 
In this paper, we have presented a statistical approach which works towards this.
We studied several statistical tools to derive risk prediction and tolerance bounds in the context of nuclear waste characterization.
The main challenge was related to the small number of data which are usually available in real-world situations.
In this context, the normality assumption is  generally unfounded, especially in the case of strongly asymmetrical data distributions, which are common in real-world characterization studies.
Much narrower bounds exist in the statistical literature, and this paper has highlighted them.
Moreover, these are distribution-free tools and no strong assumptions are needed, e.g., with respect to the normality of the distribution of the  variable under consideration.
These tools are distribution statistics aids which can provide practical confidence bounds for radiological probabilistic risk assessment.

Certain concentration inequalities, used in a conservative way (with a boostrapping technique), have shown to be strongly robust.
However, the prediction and tolerance bounds given by the standard Bienaym\'e-Chebychef inequality are very loose. Thus, their use in risk assessment leads to unnecessarily high  conservatism.
If their assumptions (unimodality and tail convexity of the pdf) can be justified, the Camp-Meidell and Van Dantzig inequalities should be considered first.
In the absence of any assumptions,  Wilks' formula offers the advantage of directly giving an upper bound of the risk of being non-conservative, but is not of great advantage when dealing with very small-sized samples or low risk bounds. Indeed, in such cases, the excessive conservatism can be greater than when using the concentration inequalities. 
Moreover,  Wilks' formula can suffer from a lack of flexibility in practical situations.

In terms of future directions, more recent concentration inequalities \cite{hoeffding1963,boulug13} could be studied and may potentially give much narrower intervals.
As an aside, it has also been shown in \cite{blaioo12} how to use probabilistic inequalities to determine the precision in the estimation of the mean of a random variable from a measurement sample.
With these kinds of inequalities, we can find the minimal number of measurements required in order to reach a given confidence level in estimating the mean.
In conclusion, possible applications of these tools are numerous across all safety considerations based on expensive experimental processes.
Further research  and applied case studies could lead to the development of useful guides for practitioners, in particular in the nuclear dismantling context.

%%%%%%%%%%%%%%%%%%%%%%%%%%%%%%%%%%%
\section*{Acknowledgments}

The authors wish to thank Herv\' e Lamotte, Alexandre Le Cocguen, Dominique Carr\' e and Ingmar Pointeau from the CEA Department of Nuclear Services, and Thierry Advocat, head of the CEA GFDM research program, for allowing the use of the H$_2$ flow rates data from drums of radioactive waste.
We also thank L\'eandre Brault and Emmanuel Remy for many useful comments on this paper.
Finally, we are grateful to Kevin Bleakley for the English language corrections.

% BibTeX users please use

%

\end{document}